\documentstyle[epsf,12pt]{article}
\begin{document}

\catcode`@=11
\long\def\@caption#1[#2]#3{\par\addcontentsline{\csname
  ext@#1\endcsname}{#1}{\protect\numberline{\csname
  the#1\endcsname}{\ignorespaces #2}}\begingroup
    \small
    \@parboxrestore
    \@makecaption{\csname fnum@#1\endcsname}{\ignorespaces #3}\par
  \endgroup}
\catcode`@=12
\newcommand{\newc}{\newcommand}
\newc{\gsim}{\lower.7ex\hbox{$\;\stackrel{\textstyle>}{\sim}\;$}}
\newc{\lsim}{\lower.7ex\hbox{$\;\stackrel{\textstyle<}{\sim}\;$}}
\newc{\gev}{\,{\rm GeV}}
\newc{\mev}{\,{\rm MeV}}
\newc{\ev}{\,{\rm eV}}
\newc{\kev}{\,{\rm keV}}
\newc{\tev}{\,{\rm TeV}}
\newc{\mz}{m_Z}
\newc{\mpl}{M_{Pl}}
\newc{\chifc}{\chi_{{}_{\!F\!C}}}
\newc\order{{\cal O}}
\newc\CO{\order}
\newc\CL{{\cal L}}
\newc\CY{{\cal Y}}
\newc\CH{{\cal H}}
\newc\CM{{\cal M}}
\newc\CF{{\cal F}}
\newc\CD{{\cal D}}
\newc\CN{{\cal N}}
\newc{\eps}{\epsilon}
\newc{\re}{\mbox{Re}\,}
\newc{\im}{\mbox{Im}\,}
\newc{\invpb}{\,\mbox{pb}^{-1}}
\newc{\invfb}{\,\mbox{fb}^{-1}}
\newc{\yddiag}{{\bf D}}
\newc{\yddiagd}{{\bf D^\dagger}}
\newc{\yudiag}{{\bf U}}
\newc{\yudiagd}{{\bf U^\dagger}}
\newc{\yd}{{\bf Y_D}}
\newc{\ydd}{{\bf Y_D^\dagger}}
\newc{\yu}{{\bf Y_U}}
\newc{\yud}{{\bf Y_U^\dagger}}
\newc{\ckm}{{\bf V}}
\newc{\ckmd}{{\bf V^\dagger}}
\newc{\ckmz}{{\bf V^0}}
\newc{\ckmzd}{{\bf V^{0\dagger}}}
\newc{\X}{{\bf X}}
\newc{\bbbar}{B^0-\bar B^0}
\def\bra#1{\left\langle #1 \right|}
\def\ket#1{\left| #1 \right\rangle}
\newc{\sgn}{\mbox{sgn}\,}
\newc{\m}{{\bf m}}
\newc{\msusy}{M_{\rm SUSY}}
\newc{\munif}{M_{\rm unif}}
%
%
\def\NPB#1#2#3{Nucl. Phys. {\bf B#1} (19#2) #3}
\def\PLB#1#2#3{Phys. Lett. {\bf B#1} (19#2) #3}
\def\PLBold#1#2#3{Phys. Lett. {\bf#1B} (19#2) #3}
\def\PRD#1#2#3{Phys. Rev. {\bf D#1} (19#2) #3}
\def\PRL#1#2#3{Phys. Rev. Lett. {\bf#1} (19#2) #3}
\def\PRT#1#2#3{Phys. Rep. {\bf#1} (19#2) #3}
\def\ARAA#1#2#3{Ann. Rev. Astron. Astrophys. {\bf#1} (19#2) #3}
\def\ARNP#1#2#3{Ann. Rev. Nucl. Part. Sci. {\bf#1} (19#2) #3}
\def\MPL#1#2#3{Mod. Phys. Lett. {\bf #1} (19#2) #3}
\def\ZPC#1#2#3{Zeit. f\"ur Physik {\bf C#1} (19#2) #3}
\def\APJ#1#2#3{Ap. J. {\bf #1} (19#2) #3}
\def\AP#1#2#3{{Ann. Phys. } {\bf #1} (19#2) #3}
\def\RMP#1#2#3{{Rev. Mod. Phys. } {\bf #1} (19#2) #3}
\def\CMP#1#2#3{{Comm. Math. Phys. } {\bf #1} (19#2) #3}
\relax
%
%
%
\def\beq{\begin{equation}}
\def\eeq{\end{equation}}
\def\bea{\begin{eqnarray}}
\def\eea{\end{eqnarray}}
%
%
%
\newc{\ie}{{\it i.e.}}          \newc{\etal}{{\it et al.}}
\newc{\eg}{{\it e.g.}}          \newc{\etc}{{\it etc.}}
\newc{\cf}{{\it c.f.}}
%
%
%
%
\def\slash#1{\rlap{$#1$}/} 
\def\Dsl{\,\raise.15ex\hbox{/}\mkern-13.5mu D} 
\def\delsl{\raise.15ex\hbox{/}\kern-.57em\partial}
\def\Ksl{\hbox{/\kern-.6000em\rm K}}
\def\Asl{\hbox{/\kern-.6500em \rm A}}
\def\Qsl{\hbox{/\kern-.6000em\rm Q}}
\def\gradsl{\hbox{/\kern-.6500em$\nabla$}}
%
%
%
\def\bar#1{\overline{#1}}
\def\vev#1{\left\langle #1 \right\rangle}
%

\begin{titlepage}
\begin{flushright}
OSU--HEP--99--10\\
LBNL--44284\\
UCB--PTH--99/43\\
hep-ph/9909476\\
September 1999\\
\end{flushright}
\vskip 2cm
\begin{center}
{\large\bf Higgs--Mediated $B^0\to\mu^+\mu^-$ 
in Minimal Supersymmetry}
\vskip 1cm
{\normalsize\bf
K.S.\ Babu$\,{}^1$ and Christopher Kolda$\,{}^{2,3}$} \\
\vskip 0.5cm
{\it ${}^1\,$Department of Physics, Oklahoma State University\\
Stillwater, OK~~74078, USA\\ [0.1truecm]
${}^2\,$Theory Group, MS 50A-5101, Lawrence Berkeley National Laboratory\\
1 Cyclotron Road, Berkeley, CA~~94720, USA\\[0.1truecm]
${}^3\,$Department of Physics, University of California, Berkeley, 
CA~~94720, USA\\[0.1truecm]
}

\end{center}
\vskip .5cm

\begin{abstract}
In this letter we demonstrate a new source for large flavor--changing
neutral currents within the minimal supersymmetric standard model. At
moderate to large $\tan\beta$, it is no longer possible to diagonalize 
the masses of the quarks in the same basis as their Yukawa couplings.
This generates large flavor--violating couplings of the form 
$\bar b_Rd_L \phi$ and $\bar b_Rs_L \phi$ where $\phi$ is any of the 
three neutral, physical Higgs bosons.  These new couplings lead to rare 
processes in the $B$ system such as $B^0 \to\mu^+ \mu^-$
decay and $\bbbar$ mixing. We show that the latter is 
anomalously suppressed, while the former is in the experimentally
interesting range.  Current limits on $B^0 \to \mu^+ \mu^-$ already
provide nontrivial constraints on models of moderate to large 
$\tan\beta$, with an observable signal possible at Run II of the 
Tevatron if $m_A \lsim 400-700$ GeV, extending to the TeV range if a
proposed Run III of $30\invfb$ were to occur.
\end{abstract}

\end{titlepage}

\setcounter{footnote}{0}
\setcounter{page}{1}
\setcounter{section}{0}
\setcounter{subsection}{0}
\setcounter{subsubsection}{0}


Extensions of the Standard Model containing more than one Higgs SU(2)
doublet generically allow flavor-violating couplings of the
neutral Higgs bosons. Such couplings, if unsuppressed, will lead to 
large flavor-changing
neutral currents, in direct opposition to experiment~\cite{ars}.
Models such as the Minimal Supersymmetric Standard Model (MSSM) avoid
these dangerous couplings by segregating the quark and Higgs fields so 
that one Higgs ($H_u$) can couple only to $u$-type quarks while the
other ($H_d$) couples only to $d$-type. Within unbroken supersymmetry
this division is completely natural; in fact, it is required
by the holomorphy of the superpotential.

However, after supersymmetry is broken, there is nothing left to
protect this division. In fact, it has been known for some time that 
couplings of the form $QU^cH_d^*$ and $QD^cH_u^*$ are generated at
one-loop~\cite{old}. As such, one would expect some flavor violation
to arise in the neutral Higgs sector, but always suppressed by loops
and therefore small (as small or smaller than Standard Model
flavor-changing). But this is not the correct deduction.

Hall, Rattazzi and Sarid (HRS)~\cite{hrs} showed that at moderate to
large $\tan\beta\equiv\vev{H_u}/\vev{H_d}$ 
the contributions to $d$-quark masses coming from
the non-holomorphic operator $QD^cH_u^*$ can be equal in size to those
coming from the usual holomorphic operator $QD^cH_d$ despite the loop
suppression suffered by the former. This is because the operator
itself gets an additional enhancement of $\tan\beta$. That is, the product 
$\tan\beta/16\pi^2$ need not be very small as $\tan\beta$ approaches
its upper bound of 60 to 70.

The HRS result was followed shortly by Ref.~\cite{brp} 
which analyzed the entire $d$-quark mass
matrix in the presence of these corrections and found appreciable
contributions to the CKM mixing angles. 
It has also recently been realized that the HRS corrections can
significantly alter the (flavor-conserving) couplings of the Higgs 
bosons~\cite{cmw,bk}. In this letter we take our
analysis from Ref.~\cite{bk} one step further 
and show that flavor-changing couplings of the neutral Higgs
bosons are also generated. 
We will show that these couplings can be appreciable and can
be so even without invoking squark mixing and/or non-minimal K\"ahler 
potentials~\cite{hpt}, and remain large even in the limit of heavy
squarks and gauginos. These new couplings
generate a variety of flavor-changing processes, including $\bar
B^0-B^0$ mixing and decays such as $B^0\to\mu^+\mu^-$ which we will
study in this letter. A more complete
discussion of these and other effects will be found in a forthcoming
paper~\cite{coming}. 

We begin by writing the effective Lagrangian for the interactions of
the two Higgs doublets with the quarks in an arbitrary basis:
\beq
-\CL_{ef\!f}=\overline{D}_R\yd Q_L H_d 
+ \overline{D}_R\yd \left[\eps_g+\eps_u \yud\yu\right] Q_L H_u^*+h.c.
\label{leff1}
\eeq
Here $\yd$ and $\yu$ are the $3\times3$ Yukawa matrices of the
microscopic theory, while 
the $\eps_{g,u}$ are the finite, loop-generated non-holomorphic
Yukawa coupling coefficients derived by HRS. The leading contributions to
$\eps_g$ and $\eps_u$ are generated by the two diagrams in
Fig.~1. 
(There can also be contributions to $\CL_{ef\!f}$ proportional to
$\yd\ydd\yd$; 
however, since they are typically smaller than the $\eps_g$
contribution and do not generate flavor-violations, we will 
not consider them further.)

Consider the first diagram in Fig.~1. 
If all $\tilde Q_i$ masses are assumed degenerate at some scale
$\munif$ then, at lowest order, $i=k$ and the diagram contributes only to 
$\eps_g$:
\beq
\eps_g \simeq\frac{2\alpha_3}{3\pi}\mu^*M_3 f(M_3^2,m^2_{\tilde
  Q_L},m^2_{\tilde d_R}), 
\eeq
where~\cite{hrs}
\beq
f(x,y,z)=-\frac{xy\log(x/y)+yz\log(y/z)+zx\log(z/x)}{(x-y)(y-z)(z-x)}.
\eeq
Meanwhile, the second diagram of Fig.~1 contributes to $\eps_u$:
\beq
\eps_u \simeq \frac{1}{16\pi^2}\mu^* A_U f(\mu^2,m^2_{\tilde Q_L},m^2_{\tilde 
  u_R}).
\label{epsu1}
\eeq
(We assume that the trilinear $A$-terms
can be written as some flavor-independent mass times $\yu$.)
For typical inputs, one usually finds $|\eps_g|$ is about 4 times larger
than $|\eps_u|$.

However, there is another sizable contribution to $\eps_u$, this one 
coming from the {\em first}\/ diagram in Fig.~1.
It is well-known that $\tilde Q_i$ degeneracy is broken by
radiative effects induced by Yukawa couplings. While this would appear 
to be a higher-order effect, for $\munif\gg \msusy$ it is
amplified by a large logarithm and thus can be $\CO(1)$. Without
resumming that log, one finds a deviation from universality of~\cite{dgh}
\beq
\Delta\m^2_{\tilde Q}\simeq -\frac{1}{8\pi^2}(3m_0^2+A_0^2)
\left[\yud\yu+\ydd\yd\right]\log\left(\frac{\munif}{\msusy}\right)
\eeq
where $m_0$ and $A_0$ are the common scalar mass and trilinear soft
term at $\munif$. 
(Resummation can bring in additional flavor structure such as 
$(\yud\yu)^2$, $(\ydd\yd)^2$, $(\yud\yu)(\ydd\yd)$, etc.,
but these are numerically less significant and do not lead to any new
flavor structure.)
At the SUSY scale, we can write the $\tilde Q$ mass matrix in the form 
\beq
\m^2_{\tilde Q}=\bar m^2 \left({\bf 1} + c \yud\yu + c\ydd\yd\right)
\label{nonuniv}
\eeq
where 
\beq
c\simeq-\frac{1}{8\pi^2}\frac{3m_0^2+A_0^2}{\bar m^2}
\log\left(\frac{\munif}{\msusy}\right) 
\eeq
and $\bar m^2$ is a flavor-independent mass term.
The effect of this non-universality is to
generate a contribution to $\eps_u$ proportional to $\alpha_3$ and
thus potentially large (the $\ydd\yd$ piece is again irrelevant).
Specifically,
\beq
\Delta\eps_u\simeq\left\lbrace\begin{array}{l}
-c\eps_g/3 \quad (m^2_{\tilde Q}\simeq M^2_3) \\
-c\eps_g/2 \quad (m^2_{\tilde Q}\gg M^2_3)
\end{array} \right. .
\label{epsu2}
\eeq
If $\munif$ is identified as the GUT scale, then $c$ is typically in 
the range $-1\lsim c\lsim-\frac14$. Thus, this second contribution can either
dramatically increase $\eps_u$ or potentially cancel much of it off,
depending on their relative (model-dependent) signs. Perhaps more
importantly, this contribution can still lead to large $\eps_u$ even
if the $A$-terms at the weak scale are small compared to
the squark masses.

Now we return to Eq.~(\ref{leff1}). We can simplify it 
considerably by working in a basis in
which $\yu=\yudiag$ and $\yd=\yddiag\ckmzd$ where $\ckmz$ is the CKM
matrix at lowest-order (the meaning of this will be clear
shortly) and $\yudiag$ and $\yddiag$ are both diagonal. Then
\beq
-\CL_{ef\!f}=\overline D_R\,\yddiag\ckmzd\, Q_L H_d 
+ \overline D_R\,\yddiag\ckmzd \left[\eps_g
+\eps_u \yudiagd\yudiag\right] Q_L H_u^*+h.c.
\label{leff2}
\eeq
It is clear
that in the absence of the $\eps_u$ term, all pieces of the effective
Lagrangian can be diagonalized in the same basis, preventing the
appearance of flavor-changing neutral currents (FCNCs). 
It is the presence of the $\eps_u\yudiagd\yudiag$ 
piece, however, that will prevent simultaneous diagonalization and generate
some flavor-changing. 

To see how this works, it is sufficient to keep only the
Yukawa couplings of the third generation so that
$(\yudiag)_{ij}=y_t\delta_{i3}\delta_{j3}$ and
$(\yddiag)_{ij}=y_b\delta_{i3}\delta_{j3}$. 
The flavor-conserving 
pieces of $\CL_{ef\!f}$ then have the form
\beq
\left(1+\eps_g\right)\yddiag\ckmzd
=(1+\eps_g)y_b
\left(\begin{array}{ccc} 0 & 0 & 0 \\ 0 & 0 & 0 \\ 
V^0_{ub} & V^0_{cb} & V^0_{tb} \end{array}\right)
\eeq
while the flavor-changing piece has the form
\beq
\eps_u\yddiag\ckmzd\yudiagd\yudiag=\eps_u y_t^{2}y_b
\left(\begin{array}{ccc} 
0 & 0 & 0 \\ 0 & 0 & 0 \\ 0 & 0 & V^0_{tb} \end{array}\right).
\eeq
We can define a physical eigenbasis by rotating the $d$-component of 
$Q_L$ by a new matrix $\ckm$ defined by diagonalizing the mass
matrix:
\beq
\left(\ckmd\CY^\dagger\CY\ckm\right)_{ij}={\rm diag}(\bar y_d^2,\bar
y_s^2, \bar y_b^2)
\eeq
where the $\bar y_i$ are the defined to be the ``physical'' Yukawa 
couplings, \eg,
$m_b=\bar y_b v_d$; and 
\beq
\CY=\yddiag\ckmzd\left[1+\tan\beta\left(\eps_g
+\eps_u\yudiagd\yudiag\right)\right],
\eeq
the $\tan\beta$ coming from the vev of $H_u$ which multiplies the
loop-induced terms. $\ckm$ can now be interpreted as the physical CKM
matrix. 

In the physical basis, the (3,3) element of the mass matrix gives us
the corrected $b$-quark mass:
\beq
\bar y_b\simeq y_b \left[1+(\eps_g 
+\eps_u y_t^2)\tan\beta\right].
\label{yb}
\eeq
To get to this equation we used the fact that
one finds no large (\ie, $\tan\beta$-enhanced) corrections to
$V_{tb}$~\cite{brp}, so that we can replace 
$V^0_{tb}\simeq V_{tb}\simeq 1$.

The corrected CKM elements are the elements of $\ckm$. In particular,
\beq
V_{ub}\simeq V^0_{ub}\left[\frac{1+\eps_g\tan\beta}
{1+(\eps_g+\eps_u y_t^2)\tan\beta}\right].
\label{vub}
\eeq
The same form also holds for the corrected $V_{cb}$, $V_{td}$ and $V_{ts}$.
Consistent with our earlier simplification that
$V^0_{tb}\simeq1$, one finds that $V_{tb}$ receives no correction.
Note that Eqs.~(\ref{yb})--(\ref{vub}) present a coherent picture of the
radiative corrections generated by SUSY-breaking: all of the diagrams
represented in Fig.~1 contribute to a renormalization of the mass, but only the
higgsino-mediated diagrams contribute a piece to the mass matrix which is not
diagonal in the usual mass basis and which therefore generates FCNCs. Thus we
see that $V_{ub}$ reduces to $V^0_{ub}$ in the limit that $\eps_u=0$.

For $\eps_u\neq0$, however, the rotation that diagonalized the mass matrix does
not diagonalize the Yukawa couplings of the Higgs fields. Redefining $D_L$ and
$D_R$ as the mass eigenstates, 
the effective Lagrangian for their couplings to the neutral Higgs
fields is
\beq
-\CL_{d,ef\!f}=\bar D_R \,\yddiag\ckmzd\ckm\, D_L H_d^0+
\bar D_R\,\yddiag\ckmzd\left[\eps_g +\eps_u\yudiagd\yudiag\right]\ckm 
\, D_L H_u^{0*}+h.c.
\eeq
Keeping only the flavor changing pieces, this simplifies after some
algebra to
\beq
\CL_{{}_{F\!C\!N\!C}}=\frac{\bar y_bV^*_{tb}}{\sin\beta}\,
\chifc
\left[V_{td}\bar b_Rd_L+V_{ts}\bar b_Rs_L\right]
\left(\cos\beta H_u^{0*}-\sin\beta H_d^0\right) +h.c.
\label{final}
\eeq
with the quark fields in the physical/mass eigenbasis,
and defining
\beq
\chifc=\frac{-\eps_uy_t^2\tan\beta}{(1+\eps_g\tan\beta)
[1+(\eps_g+\eps_uy_t^2)\tan\beta]}
\eeq
to parameterize the amount of flavor-changing induced.
Note also that we have expressed the coupling in terms of $\bar
y_b=m_b/v_d$ instead of the original, but unphysical, $y_b$. 

The final step is to define the neutral Higgs mass eigenstates. These are
defined as usual:
\bea
h^0&=&\sqrt{2}(\cos\alpha \,\re H^0_u-\sin\alpha \,\re H^0_d),\nonumber\\
A^0&=&\sqrt{2}(\cos\beta \,\im H^0_u+\sin\beta \,\im H^0_d)
\eea
and $H^0$ orthogonal to $h^0$. Then the flavor-changing couplings between the
Higgs mass states and the fermion mass states are:
\beq 
\left.
\begin{array}{rc}
h^0\bar b_R d_L:& i\cos(\beta-\alpha) \\
H^0\bar b_R d_L:& i\sin(\beta-\alpha) \\
A^0\bar b_R d_L:& 1
\end{array} \right\rbrace
\times \frac{\bar y_b V_{td} V^*_{tb}}{\sqrt2\sin\beta}\,\chifc
\eeq
A similar expression holds for the Higgs couplings to $\bar b_R s_L$ with
$V_{td}$ replaced by $V_{ts}$.
One non-trivial check of this result is to take the 
Higgs decoupling limit in which 
$m_{A^0}\to \infty$, driving $\alpha\to\beta-\frac\pi2$. There the
$h^0\bar b_Rd_L$ coupling goes to zero as it should in any single Higgs
doublet model.

We will now consider two processes which constrain and/or provide a
signal for the Higgs-mediated FCNCs: $\bbbar$ mixing and the decay
$B^0\to\mu^+\mu^-$. The case of $\bbbar$ mixing is actually quite
amusing. $\Delta m_{B_d}$ is very well known and usually provides one of
the tightest constraints on new sources of flavor-violation in the
$d$-quark sector. And, in principle, mixing can be generated by 
single Higgs exchange.
The leading order contribution of the 3 physical Higgs bosons to an
effective operator $\bar b_R^i d_L^i\bar b_R^j d_L^j$  ($i,j$ are
SU(3) indices) is
proportional to the product of vertex factors and propagators given by:
\beq
\CF\equiv
\left[\frac{\cos^2(\beta-\alpha)}{m_h^2}
+\frac{\sin^2(\beta-\alpha)}{m_H^2}-\frac{1}{m_A^2}
\right].
\label{vanish}
\eeq 
However, $\CF=0$ at lowest order.
The existence of this zero is essentially an accidental cancellation
coming from the special form of Eq.~(\ref{final}) and not an indication
that the flavor-changing is illusory. 

It is natural to ask whether this zero survives loop corrections, and
one finds that it does not. However, the cost of adding another loop
to the diagram is high and tends to suppress this new contribution too
much to dominate the Standard Model contribution. 
We have considered in detail the largest non-zero
contribution, which arises from top-stop induced vacuum polarization on
the internal Higgs line. While these propagator corrections to the
Higgs are known to be large~\cite{haber}, we find that the
leading term (which is a correction to the $H_u$ line) is suppressed
by $1/\tan^2\beta$. The next-leading term (a correction on the $H_d$
line due to left-right stop mixing) is present but is not very
large. All other radiative corrections we expect to be even
smaller.

One can still derive a bound on $m_A$ by demanding that the MSSM
contribution to $\Delta m_{B_d}$ is less than its observed value. Such
a bound will depend sensitively on whether or not the two
contributions to $\eps_u$ from Eqs.~(\ref{epsu1}) and (\ref{epsu2}) 
interfere
constructively or destructively. Assuming all MSSM masses to be near
$500\gev$ and constructive interference, we find 
$m_A\lsim 100$ to $125\gev$ for $\tan\beta=40$ to 60.
Direct search constraints aside, it is known that
models with such a light second Higgs doublet generally 
contribute far too much to $b\to s\gamma$ and are therefore already
ruled out~\cite{bsg}. 
Thus this new source of flavor-changing rules out a part of 
parameter space which is already known to be 
disfavored. 

We now consider the rare decay $B^0\to\mu^+\mu^-$. This occurs
via emission off the quark current of a single virtual Higgs boson 
which then decays leptonically. 
The largest leptonic 
flavor-changing branching fraction would clearly be to
$\tau^+\tau^-$. However, the branching fraction to $\mu$'s is only
suppressed by $(m_\mu/m_\tau)^2$ times a phase space factor, 
which is only about 1 part in 100. The current experimental limits 
on $Br(B^0\to\mu^+\mu^-)$ are at the $10^{-6}$ level, which
means that the largest the branching ratio into $\tau$'s could be is
about $10^{-4}$. Given the extreme difficulties encountered in trying
to measure this decay experimentally, it is doubtful that the 
$\tau$-mode will ever provide an interesting constraint or signal in 
and of itself. Thus we will concentrate on the $\mu$-channel.

The amplitude for the process $B_{(d,s)}^0\to\mu^+\mu^-$ is given by:
\beq
{\cal A}=\eta_{{}_{QCD}}
\frac{\bar y_b y_\mu V_{t(d,s)}V^*_{tb}}{2 \sin\beta}\,\chifc 
\bra{0^{\phantom{0}}\!\!}\bar b_R d_L \ket{B_{(d,s)}^0} 
\left[\bar\mu\left(a_1+a_2\gamma^5\right)\mu\right]
\eeq
where
\bea
a_1&=&\frac{\sin(\beta-\alpha)\cos\alpha}{m_H^2} - 
\frac{\cos(\beta-\alpha)\sin\alpha}{m_h^2},\nonumber\\
a_2&=&-\frac{\sin\beta}{m_A^2}.
\eea
The partial width is then
\beq
\Gamma(B^0_{(d,s)}\to\mu^+\mu^-)=\frac{\eta^2_{{}_{QCD}}}{128\pi}\,
m_B^3 f_B^2\,
\bar y_b^2 y_\mu^2\, |V_{t(d,s)}^* V_{tb}|^2\,\chifc^2 (a_1^2+a_2^2).
\label{width}
\eeq
In the large $m_A$, large $\tan\beta$ limit, $a_1^2+a_2^2\simeq 2/m_A^4$.
The QCD correction is identical to the usual running of a quark mass 
operator, 
which in this case gives $\eta_{{}_{QCD}}$ between 1.4 and 1.6 for 
$m_A$ between $m_Z$ and $500\gev$.
Experimentally, $Br(B^0_{(d,s)}\to\mu^+\mu^-)<(6.8,20)\times10^{-7}$
at 90\% confidence~\cite{cdf}. 
Thus $\Gamma_{(d,s)}<(2.9,8.7)\times10^{-19}\gev$.
The factor of 3 in going from the $B_d^0$ to the $B_s^0$ limits is due to
a factor of 3 suppression in the production cross-section of $B_s^0$ 
compared to $B_d^0$ at the Tevatron. However, theory predicts the partial 
width for $B_s^0\to\mu^+\mu^-$ 
to be enhanced by $(V_{ts}/V_{td})^2\simeq 25$. Thus 
one expects a signal in $B_s^0$ decays before one is observed in $B_d^0$.

A few quick estimates can give us an impression of the importance of
these new contributions. For nearly-degenerate MSSM particles at
$500\gev$, one finds $|\eps_g|\approx 1/80$ and $|\eps_u|\approx
(1/4)|\eps_g|$, not including  in $\eps_u$ the contribution 
of Eq.~(\ref{epsu2}). We derive a bound on $m_A$  
from the limit on $B_s^0\to\mu^+\mu^-$ and using $f_B=180\mev$
and $|V_{ts}|=0.04$. The bound depends sensitively on the  
signs of $\eps_g$ and $\eps_u$ as well as the size of the
$c$-parameter of Eq.~(\ref{epsu2}), which we take in the range
$-3/4\leq c\leq0$. We also demand that $y_b\leq y_t$ to avoid problems 
with perturbation theory and consistency with unification; this places 
an upper bound on $\tan\beta$ as a function of $\eps_g$, $\eps_u$ and $c$.
Varying over all of these, the strongest bounds are
\beq
m_A>(225,175,230,215)\gev 
\eeq
for $\tan\beta= (29,65,38,65)$, 
$c=(-3/4,0,0,-3/4)$ and the signs of $\{\eps_g,\eps_u\}$ being 
$(--,+-,-+,++)$ respectively.
 
Like the case of $\bbbar$ mixing, we are finding ourselves
in the range already constrained by $b\to s\gamma$ and direct searches.
However, unlike the mixing case where the MSSM contribution
was typically smaller than the Standard Model prediction, here we are
still far above the Standard Model which predicts
$Br(B^0_{(d,s)}\to\mu^+\mu^-)\simeq
(1.5,35)\times10^{-10}$~\cite{sm}. Thus
further experimental data can significantly improve the bounds on
$m_A$ or find a non-zero signal induced by supersymmetry.

So what is implied for Run~II at the Tevatron? Assuming no
change in their efficiencies and acceptances, CDF can in principle place
a bound $Br(B^0_s\to\mu^+\mu^-)<1\times10^{-7}$ given $1\invfb$ of
data, a factor of 20 stronger than present. Thus the region probed in
$m_A$ will increase by $20^{1/4}\simeq2$:
\beq
m_A>(475,365,490,450)\gev
\eeq
for the same sets of inputs as previously. After collecting $5\invfb$
these masses increase by another 50\%, up to $725\gev$.
Finally, if the proposed ``Run III'' of the Tevatron with $30\invfb$
were to occur, masses of $A^0$ all the way to $1\tev$ could be studied. 
This could be a very important signal for supersymmetry since
{\em this source of flavor-changing does not decouple as
$\msusy\to\infty$}\/ so long as $m_A$ does not also get very heavy. 
That is to say, the
bound on $m_A$ is roughly independent of $\msusy$. Therefore
supersymmetric spectra in the multi-hundred GeV to TeV range may
be probed at the Tevatron through rare $B$-decays even when direct
production of supersymmetry (including the second Higgs doublet)
cannot be observed.
Since the precise predictions for $Br(B^0_s\to\mu^+\mu^-)$ are highly
dependent on the individual model, these estimates should only be
taken as indicative. Further work will be forthcoming~\cite{coming}.

It is also possible to look for new sources of flavor-changing in
inclusive semileptonic decays $B\to X_s\mu^+\mu^-$. The width for this 
process can be extracted from Eq.~(\ref{width}) by replacement of
$f_B$ with $m_B$ and dividing by $192\pi^2$ for the 3-body phase
space. The rate is thus a factor of 10 smaller than for 
$B_s -> \mu+ \mu-$.
Comparing to current bounds~\cite{cleo} yields constraints on
$m_A$ that are weaker by a factor of 1.8
than the bounds from the purely leptonic
mode. The ability of future experiments to extract information from
this mode will be discussed in \cite{coming}.

Finally, we find it noteworthy that the largest signals tend to occur
for $\eps_g<0$ and intermediate values of $\tan\beta$. 
In minimal GUT models, one expects unification of the
$b$- and $\tau$-Yukawa couplings. But it is well-known that this
unification fails over most of the parameter space of the MSSM and
generally necessitates the use of the HRS corrections to bring the
Yukawas back into agreement. Typically one requires
$(\eps_g+y_t^2\eps_u)\tan\beta\approx-0.2$~\cite{hrs,copw} which in
turn means that $\eps_g<0$. This provides an argument for believing
that the signal might lie in the observable range, as well as providing
another test of Yukawa unification (beyond those discussed in \cite{bk} for
flavor-conserving processes).

In summary, we have found that neutral Higgs bosons are capable of
mediating flavor-changing interactions within the MSSM. This result is
generic and does not rely on assumptions about sparticle mass 
non-universality which are usually required in order to get FCNCs. These
interactions are enhanced at large $\tan\beta$ and are in
the range that will be experimentally probed in the near future.

~

{\em Acknowledgements.}~We would like to thank 
R.~Cahn, L.~Hall, I.~Hinchliffe, A.~Kagan, H.~Murayama, M.~Suzuki
and M.~Worah for helpful conversations. KSB would also like to thank
the Theory Group at Lawrence Berkeley National Laboratory for their
hospitality during his stay when this work
was started. This work was supported in
part by the Department of Energy under contracts DE--AC03--76SF00098
and FG03--98ER41076 and by funds provided by Oklahoma State University.

~


\begin{figure}[h]
\centering
\epsfysize=1.25in
\hspace*{0in}
\epsffile{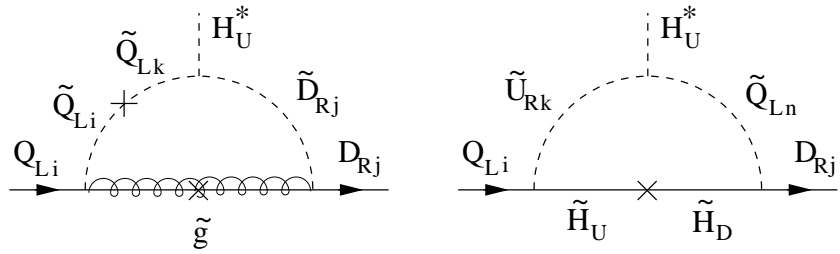}
\caption{Leading contributions to $\eps_g$ and $\eps_u$. Indices
  $i,j,k,n$ label flavors.}
\label{fig:eps}
\end{figure}

\end{document}